\documentclass[12pt,preprint]{aastex}

\begin{document}

\title{Triggering Collapse of the Presolar Dense Cloud Core 
and Injecting Short-Lived Radioisotopes with a Shock Wave. 
III. Rotating Three Dimensional Cloud Cores}

\author{Alan P.~Boss and Sandra A.~Keiser}
\affil{Department of Terrestrial Magnetism, Carnegie Institution for
Science, 5241 Broad Branch Road, NW, Washington, DC 20015-1305}
\email{boss@dtm.ciw.edu}

\begin{abstract}

 A key test of the supernova triggering and injection hypothesis
for the origin of the solar system's short-lived radioisotopes is to 
reproduce the inferred initial abundances of these isotopes. We present
here the most detailed models to date of the shock wave triggering
and injection process, where shock waves with varied properties strike
fully three dimensional, rotating, dense cloud cores. The models are
calculated with the FLASH adaptive mesh hydrodynamics code. Three 
different outcomes can result: triggered collapse leading to fragmentation 
into a multiple protostar system; triggered collapse leading to a single 
protostar embedded in a protostellar disk; or failure to undergo
dynamic collapse. Shock wave material is injected into the collapsing
clouds through Rayleigh-Taylor fingers, resulting in initially inhomogeneous 
distributions in the protostars and protostellar disks. Cloud rotation
about an axis aligned with the shock propagation direction does not increase 
the injection efficiency appreciably, as the shock parameters were chosen 
to be optimal for injection even in the absence of rotation. For a shock wave 
from a core-collapse supernova, the dilution factors for supernova material 
are in the range of $\sim 10^{-4}$ to $\sim 3 \times 10^{-4}$, in 
agreement with recent laboratory estimates of the required amount of 
dilution for $^{60}$Fe and $^{26}$Al. We conclude that a type II supernova
remains as a promising candidate for synthesizing the solar system's 
short-lived radioisotopes shortly before their injection into the presolar
cloud core by the supernova's remnant shock wave.

\end{abstract}


\section{Introduction}

 Primitive meteorites offer a direct link between the production of
heavy elements in stellar interiors and explosions and the incorporation
of these elements into the pebbles and planetesimals that formed the
planets of our solar system. Chondritic meteorites contain presolar
grains, with exotic isotopic ratios indicative of their stellar origins,
as well as cm-size refractory inclusions, thought to be the oldest surviving 
solids formed in the hottest regions of the solar nebula, based on their 
abundances of the decay products of short- and long-lived radioactive isotopes.
An ongoing challenge is to use this rich meteoritical record to discern
how the solar system came to be formed from the debris ejected by previous
generations of wind-emitting and explosive stars (e.g., MacPherson \&
Boss 2011).

 $^{60}$Fe requires nucleosynthesis in a high mass asymptotic giant branch 
(AGB) star or a type II supernova (SNe) for its production in significant 
quantities (e.g., Mishra \& Goswami 2014). The half-life of $^{60}$Fe has
been redetermined to be $2.62 \pm 0.04$ Myr (Rugel et al. 2009). Evidence 
for the presence of live $^{60}$Fe during the formation of chondrites (e.g., 
Tachibana et al. 2006; Dauphas \& Chaussidon 2011) thus has been seen as the 
strongest argument in favor of the injection of $^{60}$Fe and other short-lived 
radioisotopes (SLRIs) into the presolar cloud (Boss 1995; Gritschneider et al.
2012) or the solar nebula (Ouellette et al. 2007, 2010) by a shock wave propagating 
from a massive star that synthesized the SLRIs. In support of the former 
(cloud) scenario, the W44 type II supernova remnant (SNR) is observed to be striking 
the W44 giant molecular cloud (GMC), within which are embedded molecular cloud cores 
that could be triggered into collapse by the W44 shock front (Sashida et al. 
2013). However, cosmochemical support for either of these two scenarios has been 
in flux lately. Some laboratory work has lowered the inferred initial abundances 
of $^{60}$Fe (Moynier et al. 2011; Telus et al. 2012) to values that are more 
consistent with the galactic background abundance, rather than a nearby supernova 
or AGB star (Tang \& Dauphas 2012). This explanation might require the high levels 
of initial $^{26}$Al found in chondrites to be derived from a pre-SNe Wolf-Rayet (WR) 
star wind, which is expected to be rich in $^{26}$Al and poor in $^{60}$Fe 
(Tang \& Dauphas 2012). On the other hand, most recently Mishra \& Goswami (2014) 
used correlated initial $^{60}$Fe and $^{26}$Al abundances in chondrules from 
primitive chondrites to infer an initial $^{60}$Fe abundance for the solar nebula
similar to that originally claimed (e.g., Tachibana et al. 2006). Hence the $^{60}$Fe
evidence for a SNe or AGB origin seems to have now come full circle. Scenarios have 
also been advanced for accounting for the inferred levels of both $^{60}$Fe and 
$^{26}$Al through supernova injection into GMCs (Pan et al. 2012; Gounelle \& Meynet 
2012; Vasileiadis et al. 2013), although abundance problems remain.

 The presence of live $^{26}$Al in refractory inclusions (e.g., calcium, 
aluminum-rich inclusions -- CAIs) was the original motivation for the SNe trigger 
hypothesis (Cameron \& Truran 1977). The fact that the FUN (fractionation unknown 
nuclear) refractory inclusions show no evidence for live $^{26}$Al, coupled with the
significant $^{26}$Al depletions found in some CAIs and refractory grains,
implies that these refractory objects may have formed prior to the injection, 
mixing, and transport of $^{26}$Al into the refractories-forming region of the 
solar nebula (Sahijpal \& Goswami 1998; Krot et al. 2012; Kita et al. 2013;
Mishra \& Chaussidon 2014). The $^{26}$Al data alone, therefore, seems to
require the late arrival of SLRIs derived from a SNe into the inner region 
of the solar nebula, as opposed to injection into a giant molecular cloud complex,
followed by thorough mixing and SLRI homogenization prior to the collapse of the 
presolar cloud core.

 Detailed adaptive mesh refinement (AMR) hydrodynamical modeling has shown that 
SNe shock waves are preferred as a means for simultaneously achieving triggered 
collapse of the presolar cloud and injection of SLRIs carried by the shock wave 
(Boss \& Keiser 2013, hereafter BK13). Planetary nebula shock waves from AGB stars 
have thicknesses of $\sim 0.1 - 0.2$ pc (Jacoby et al. 2001; Pierce et al. 2004): 
BK13 found that such shocks were too thick to simultaneously trigger collapse and
result in significant injection. WR star winds were found to be likely to 
shred cloud cores, rather than induce collapse (BK13). These studies also showed 
that injection into a {\it rotating} cloud can increase injection efficiencies by 
as much as a factor of 10. However, these models (BK13) were restricted to axisymmetry 
about the target cloud's rotation axis (i.e., two-dimensional -- 2D). Boss \& Keiser 
(2012, hereafter BK12) presented the first 3D AMR hydrodynamic calculations of the 
shock wave triggering and injection process, using the 3D Cartesian coordinate 
version of the FLASH2.5 AMR code. However, these 3D clouds were not assumed to be 
rotating. The pioneering calculations by Boss (1995) included rotating clouds, but 
the numerical code used contracting spherical coordinates and was not a fully AMR 
code. Here we extend the 3D AMR modeling effort to include rotating cloud cores,
allowing protoplanetary disks to form along with the central protostellar objects. 
We wish to learn what effect these more realistic hydrodynamic models might have
on SLRI injection efficiencies and hence on the SNe trigger and injection hypothesis.

\section{Initial Conditions}

 The rotating presolar cloud cores are similar to those previously studied in 2D (BK13):
2.2 $M_\odot$, Bonnor-Ebert spheres with radii of 0.053 pc, rotating with angular 
velocities of either $\Omega_c = 10^{-14}$ or $10^{-13}$ rad s$^{-1}$, typical of observed
dense cloud cores (Goodman et al. 1993; Harsono et al. 2014). Even without
rotation, these cloud cores are stable against collapse for at least $\sim$ 1 Myr
(Boss et al. 2010), indicative of their marginal stability. The shock waves propagate 
along the cloud's rotation axis at speeds of $v_s = 20$, 40, or 60 km s$^{-1}$, with shock 
widths of $w_s = 3 \times 10^{-3}$ or $3 \times 10^{-4}$ pc, and shock densities ranging 
from $\rho_s = 3.6 \times 10^{-19}$ to $2.1 \times 10^{-17}$ g cm$^{-3}$. 
The choice of shock wave propagation along the cloud's rotation axis was made
in order to permit a direct comparison with the 2D models of BK13;
in 2D there is no other choice possible for this orientation. [3D models with cloud
rotation axes perpendicular to the shock propagation direction are currently 
underway and will be the subject of the next paper in this series.]

 Further details about these parameter choices and their relevance to SNe shock
waves may be found in BK12 and BK13, but we repeat to some extent, and expand 
upon, these important details here. Observations by Sashida et al. (2013) of the 
W44 type II SNR after it has been slowed down by interactions with dense clumps in the 
W44 GMC lead to estimated dense clump sizes in the range of $\sim$ 0.008 to 0.3 pc, 
a range that includes the size assumed here for the target cloud cores. Sashida et al.
(2013) also estimate expansion speeds of $\sim 13$ km s$^{-1}$ for the shocked 
molecular gas, implying significantly higher speeds earlier in the interaction process. 
In fact, Reach et al. (2005) estimated W44 SNR expansion speeds of $20-30$ km s$^{-1}$, 
closer to those assumed in our models (i.e., 20, 40, or 60 km s$^{-1}$). Boss et al. 
(2010) showed that shock speeds greater than 70 km s$^{-1}$ did not trigger 
collapse, but rather shredded the type of cloud cores studied here. 

 Chevalier (1974) showed that a SNe shock propagating into a hot ($10^4$ K) medium 
with a number density of 1 cm$^{-3}$ would slow to a top speed of $\sim 60$ km 
s$^{-1}$ within 0.25 Myr, though the top speed of the leading edge of the shock front
(the portion being modeled here) was $\sim$ 30 km s$^{-1}$. The peak number
density of the leading edge of the 
Chevalier (1974) shock was only 10 cm$^{-3}$, much lower than the
typical shock densities considered here, where $n_s = 4 \times 10^6$ cm$^{-3}$ 
for the 400 $\rho_s$ models. However, it is unclear what would happen if the 
Chevalier (1974) models were rerun for a shock propagating through a much denser, 
much colder medium, similar to the W44 GMC, or if his models included the
critical molecular cooling included here (e.g., Neufeld \& Kaufman 1993; see below). 
The W44 SNR is expanding into cold ($\sim$ 10 K) molecular gas with a number 
density $n \sim 10^2$ cm$^{-3}$ (Reach et al. 2005). BK12 noted 
that the shock density $n_s$ for an isothermal shock propagating in an ambient 
gas of density $n_{am}$ is $n_s/n_{am} = (v_s/c_{am})^2$, where $c_{am}$ is 
the ambient gas sound speed. For a model like 40-400-0.1-13, with $c_{am} = 0.2$ 
km s$^{-1}$ and $n_{am} = 10^2$ cm$^{-3}$, as in the case of the W44 SNR, the shock 
density should then be $n_s = 4 \times 10^6$ cm$^{-3}$, the same as the density 
assumed in model 40-400-0.1-13. 

 The thickness of the leading edge of the shock in the Chevalier (1974) model
is difficult to determine from his Figure 1d, as it is at the limit of the
plotted line thickness, but it appears to be less than $\sim$ 0.1 pc thick. Given
the concerns raised above, and the uncertain spatial resolution of the Chevalier 
(1974) numerical calculation, the proper shock thickness for a SNR expanding 
into a GMC like W44 is unclear. However, the Cygnus Loop is a $\sim 10^4$ yr-old  
SNR with $v_s \approx$ 170 km/sec and $w_s < 10^{15}$ cm (Blair et al. 1999).
This SNR width is consistent with our $0.1 w_s = 9.3 \times 10^{14}$ cm models.
The W44 SNR has a width less than $10^{16}$ cm (Reach et al. 2005), comparable 
to our standard shock width $w_s = 9.3 \times 10^{15}$ cm. As noted by BK13,
observed SNR appear to be most consistent with models with relatively high speed 
(70 km s$^{-1}$), thin ($0.1 w_s$) shocks, or with models with lower speed 
(10 - 40 km s$^{-1}$), thicker (1 or 0.1 $w_s$) shocks. While it is uncertain
if any realistic SNR shock has the exact parameters assumed in our models, our 
initial conditions appear to be reasonable starting points for studying the 
shock-cloud interactions in the W44 SNR, and perhaps elsewhere.

 The numerical calculations are performed with the FLASH2.5 AMR hydrodynamics
code (Fryxell et al. 2000) in Cartesian coordinates. For the numerical grid, the 
initial number of blocks is 6 along the $x$ and $z$ axes and 9 along the $y$ axis,
which is the cloud's rotation axis and the direction of propagation of the shock wave. 
Each grid block consists of $8^3$ grid points. Initially there are four levels 
of grid refinement, but the number of levels of grid refinement is increased 
to up to seven levels during the evolutions, in order to resolve the dynamically 
collapsing regions. The smallest possible grid resolution spacing reached is then
$\sim 4 \times 10^{-5}$ pc $\sim 9$ AU. A color field is used to track the evolution
of material initially in the shock front, in order to assess the degree to which
the shock front gas and dust is injected into the target cloud core. The target
cloud and ambient gas are initially isothermal at 10 K, while the shock front and
post-shock gas are initially isothermal at 1000 K. Compressional heating and cooling
by optically thin H$_2$O, CO, and H$_2$ molecules allows the thermodynamics of the 
resulting shock-cloud interaction to be followed; we used Neufeld \& Kaufman's 
(1993) radiative cooling rate of $\Lambda \approx 9 \times 10^{19} (T/100) 
\rho^2$ erg cm$^{-3}$ s$^{-1}$, where $T$ is the gas temperature in K 
and $\rho$ is the gas density in g cm$^{-3}$, for the cooling caused by 
rotational and vibrational transitions of optically thin, warm molecular gas.
Further details about our implementation of the FLASH2.5 AMR code may be 
found in Boss et al. (2010).
 
\section{Results}

 Table 1 lists the initial conditions for all of the models as well as the three
basic results: (1) collapse leading to fragmentation into a multiple protostar system,
(2) collapse leading to the formation of a single protostar in the center of a
protostellar disk, or (3) failure to trigger sustained collapse, i.e., maximum
densities no greater than $\rho \sim 10^{-16}$ g cm$^{-3}$. Most of the clouds were 
triggered into sustained collapse, reaching densities $\rho > 10^{-12}$ g cm$^{-3}$,
while some only underwent a minor contraction that did not lead to sustained collapse. 
Several of the clouds with $\Omega_c = 10^{-13}$ rad s$^{-1}$ collapsed and 
fragmented into multiple protostars. The evolutions were followed for timescales 
of $\sim 10^{5}$ yr, by which time the clouds had either collapsed, or had been pushed 
downstream by the shock wave far enough to leave the numerical grid without giving 
any indication that dynamic collapse would result.

\subsection{Protostellar Disk Formation and Fragmentation}

 Fig. 1 shows the time evolution of model 40-200-0.1-13, which serves to illustrate
the overall evolution of the models. Fig. 1a displays the initial conditions,
where the downward-moving shock front is positioned just above the top edge of
the Bonnor-Ebert-profile target cloud core. Fig. 1b shows that the shock front
crushes the edge of the cloud core and accelerates the entire core downwards.
Rayleigh-Taylor (R-T) fingers can be seen in the shock-core boundary layers, and these
R-T fingers are responsible for the injection of shock front material into the target 
cloud (e.g., BK12, BK13). By the time of Fig. 1c, the central regions of the cloud 
core have been compressed to high enough densities ($\sim 10^{-13}$ g cm$^{-3}$)
that dynamical collapse is well underway. The cooling by molecular lines throughout
these early, optically thin phases is sufficient to keep the densest regions
isothermal at $\sim$ 10 K, while the shocked regions are as hot as 1000 K. Note that
at densities above $\sim 10^{-13}$ g cm$^{-3}$, the central regions should become 
optically thick to infrared radiation, and molecular line cooling should be throttled back
considerably, but we do not attempt to include these effects in the present models;
we leave this improvement to a future set of models.
By the time of Fig. 1d, the collapsing region has begun to form a central
protostar surrounded by a large-scale, rotationally-supported protostellar
disk, with a radius of $\sim$ 500 AU.

 Fig. 2 shows the midplane of the protostellar disk for model 40-200-0.10-13
at the same time as in Fig. 1d. The protostellar disk has several distinct 
spiral arms that might undergo fragmentation during their subsequent evolution,
possibly forming secondary companions with significantly lower masses than 
that of the central protostar, which has a mass of $\sim 0.2 M_\odot$ at this 
time, with the disk mass being $\sim 0.25 M_\odot$. Fig. 3 plots the midplane
distribution of the color field, representing the SLRIs and other material
injected into the target cloud from the shock front. It can be seen that
these SLRIs are non-uniformly distributed at this early phase, as is to be
expected, given that they were injected by R-T fingers (Fig. 1b; BK12, BK13).
The SLRI abundances are higher in the regions outside the densest regions
of the protostellar disk, with the lowest abundances at the location of
the central protostar, qualitatively consistent with the idea of the late 
injection of the SLRIs to the solar nebula (e.g., Sahijpal \& Goswami 1998; 
Krot et al. 2012; Kita et al. 2013; Mishra \& Chaussidon 2014).

 Fig. 4 compares the results of two models that differ only in the
initial rotation rate of the target cloud, model 40-200-0.10-13 (Fig. 4a)
with $\Omega_c = 10^{-13}$ rad s$^{-1}$, and model 40-200-0.10-14 (Fig. 4b) 
with $\Omega_c = 10^{-14}$ rad s$^{-1}$, at roughly the same time, 0.08 Myr. 
The slower rotating model still forms a disk (Fig. 4b), but the disk
radius is only $\sim 150$ AU, compared to $\sim 500$ AU for the faster
rotating model (Fig. 4a). Fig. 5 shows that the slower rotating model
40-200-0.1-14 forms a disk without the large-scale spiral arms seen
in the faster rotating model 40-200-0.1-13. At this time the central
protostar has a mass of $\sim 0.25 M_\odot$ and the disk mass is 
$\sim 0.15 M_\odot$. Compared to model 40-200-0.1-13, then, in model
40-200-0.1-14 the protostar mass is somewhat higher, and the protostellar
disk mass is somewhat lower, as expected for a slower rotating cloud core.
Clearly both of these systems need to accrete considerably more mass
from the surrounding protostellar envelope if they are to reach the
mass of a solar-type protostar. Fig. 6 shows the SLRI abundances in
the midplane of model 40-200-0.1-14, again showing the remnant R-T
fingers and a distinct depletion in the region of the central protostar
and protostellar disk.

 Finally, Fig. 7 shows the midplane density for one of the models
that appears likely to form a multiple protostar system, model 20-200-0.1-13.
While a dominant, central protostar forms in this model, there are also
several distinct clumps evident in the spiral arms. Given the early
phase of evolution of this protostellar disk, with significant mass
yet to be accreted from the infalling envelope, these clumps may have 
a better chance of surviving the subsequent evolution and of producing a 
multiple protostar system than clumps in a protoplanetary disk, where
the most of the disk mass has already been accreted and is primarily 
draining onto the central protostar. Fig. 8 shows that the SLRI abundances
in this model are even more non-uniform than in the previous two models
discussed, as might be expected given the larger gas density variations
seen in Fig. 7, with the highest abundances of SLRIs being confined
to the regions exterior to the multiple protostellar clumps. 

 Table 1 shows that only models with the higher initial rotation rate 
considered ($10^{-13}$ rad s$^{-1}$) resulted in collapse and fragmentation, 
as is to be expected for rotationally-driven fragmentation during protostellar
collapse (e.g., Boss 1986). However, this fragmentation only occurred
for the models with the slowest shock speed considered, 20 km s$^{-1}$,
and not for otherwise identical models with 40 and 60 km s$^{-1}$ shocks,
so clearly the slower, gentler shocks were more conducive to triggering 
collapse leading to fragmentation for the faster rotating clouds.
Table 1 also shows that the models that were not compressed enough 
to undergo sustained collapse were those with the higher shock speeds
(40 and 60 km s$^{-1}$) and higher shock densities, limiting the range of
shock parameters consistent with the shock triggering scenario.
The slower rotating clouds, on the other hand, generally led to single
protostar formation, and hence remain as likely candidates for the 
presolar cloud core in this formation scenario. 

\subsection{Injection Efficiencies}

 We now turn to an estimation of the degree to which SLRIs are injected.
The injection efficiency $f_i$ is defined (BK12, BK13) as the fraction of 
the incident shock wave material that is injected into the collapsing cloud 
core. In practice, this means collapsing regions with $\rho > 10^{-16}$ 
g cm$^{-3}$. Estimated values of $f_i$ are listed in Table 1, along with 
the mass of the shock wave-derived material contained in the dynamically
collapsing regions. Note that $f_i$ was not estimated for several of 
the models in Table 1 because of the failure of a disk array that stored 
some of the model data files intended for eventual detailed analysis.

 Table 1 shows that $f_i =$ 0.032 and 0.034 for models 40-200-0.1-13 and 
40-200-0.1-14, respectively, essentially identical to the value of 
$f_i = 0.03$ estimated for the non-rotating-cloud version 40-200-0.1 (BK12) 
of these same models. Table 1's estimates of $f_i$ were obtained by calculating
the amount of color contained within the region with $\rho > 10^{-16}$ 
g cm$^{-3}$. BK12 used a somewhat different means for estimating $f_i$,
namely calculating the amount of color within $\sim 10^{16}$ cm of 
the density maximum. In order to compare the effects of these two
different estimation schemes, $f_i$ was re-estimated for 2D non-rotating
model 40-200-0.1 from BK12 using the present scheme, resulting in an estimated 
$f_i$ = 0.033, right in the middle of the values for the two new 3D models,
and close to the BK12 estimate of $f_i$ = 0.03.
However, the results of the new models differ from the expectations of the 
2D models (BK13), where it was found that including target cloud rotation 
could increase $f_i$ by a factor as large as 10. Those rotating 2D 
models studied only the so-called standard shock ($\rho_s = 3.6 \times 10^{-20}$
g cm$^{-3}$, $w_s = 3 \times 10^{-3}$ pc, $\Omega_c = 10^{-16}$ rad s$^{-1}$
through $10^{-12}$ rad s$^{-1}$), effectively 2D models 40-1-1-16 through 
40-1-1-12 in the current model notation, whereas the 3D models presented
here, 40-200-0.1-14 and 40-200-0.1-13, considered denser, thinner shocks.
In the 2D models of BK13, non-rotating model 40-1-1 led to $f_i = 0.0088$,
while non-rotating model 40-200-0.1 produced $f_i = 0.028$. This implies that
denser, thinner shocks in this range are more conducive to injection than 
the standard shock (BK13). Given that the 3D models 40-200-0.1-13 and 
40-200-0.1-14 already assumed denser, thinner shocks, the addition of 
rotation did not increase the injection efficiency appreciably further 
over that found for the non-rotating 3D model 40-200-0.1 from BK12.

 Table 1 also shows that while the estimates of $f_i$ for the models in
the middle of the table were $f_i \sim 0.03$, the two estimates at the top
of the table for models 20-10-1-14 and 20-200-0.1-14 were significantly
higher, with $f_i \sim 0.1$. This is because these two models were both
able to be evolved to considerably higher maximum densities, namely 
$\rho_{max} \sim 10^{-10}$ g cm$^{-3}$, compared to the maximum densities
obtained for models 20-600-0.1-14, 40-10-1-13, 40-10-1-14, 40-200-0.1-13,
and 40-200-0.1-14, where $\rho_{max} \sim 3 \times 10^{-12}$ g cm$^{-3}$.
This implies that if the latter five models had been able to be evolved
farther in time (i.e., if they had been recalculated with a longer $y$ grid,
in the direction of shock propagation), they would also have achieved
significantly higher injection efficiencies. Similarly, the two models
at the bottom of Table 1, 40-600-0.1-14 and 60-200-0.1-14, with 
$f_i \sim 0.006$ and 0.02, respectively, only reached $\rho_{max} 
\sim 3 \times 10^{-13}$ and $\sim 10^{-12}$ g cm$^{-3}$, 
respectively, consistent with increasing estimates for $f_i$ as
the $\rho_{max}$ achieved increases.

\section{Dilution Factors}

 The previous section shows that the injection efficiencies obtained
from these rotating 3D models are at least $f_i \sim 0.03$, and would
probably be $f_i \sim 0.1$ if higher values of $\rho_{max}$ were achieved.
We now turn to the question of what these injection efficiency estimates
imply for the abundances of SLRIs to be expected in refractory inclusions
in the SNe triggering and injection scenario.

 SNe shock waves must sweep up considerable ISM gas and dust before
striking the target cloud in order to slow down to the shock
speeds considered here and in previous models (e.g., Boss et al. 2010,
BK12, BK13) to be optimal for this scenario. Following BK12, we
define the factor $\beta$ to be the ratio of the shock front mass
originating in the SNe to the mass swept up in the intervening ISM.
With this definition, the dilution factor $D$ is defined to be the 
ratio of the amount of mass derived from the supernova to the amount of 
mass derived from the target cloud. Note that this $\beta$ factor
implicitly accounts for ``geometrical dilution'', i.e., the fact that
expanding SNRs are roughly spherical shells, rather than the planar
shocks considered here, so that as the shell expands, the column density
of SLRIs in the shell must decrease, rather then remain constant, as in 
a planar shock. However, in either case, the key point is that the
SNe-derived SLRIs will be diluted in the sense used here by the amount
of intervening ISM matter swept up as the shell expands and slows down 
by the snowplowing of the ISM. For models 40-200-0.1-13 and
40-200-0.1-14, the mass of the shock front incident on the cloud is
0.31 $M_\odot$ of gas and dust, leading to $D \approx 0.31 \beta f_i$, 
for a final system mass of $\sim$ 1 $M_\odot$. With $f_i \sim 0.03$,
then, $D \approx 0.01 \beta$. This conservative estimate assumes that the 
injected SLRIs are uniformly distributed in the resulting protostar and 
protoplanetary disk, although Figs. 3, 6, and 8 show that the protostar
is likely to be somewhat depleted in SLRIs compared to the disk, and to any
late-arriving shock wave gas and dust. 

 BK12 estimated that for a 40 km s$^{-1}$ shock model to be appropriate,
a SNe shock must have been slowed down by a factor of $\sim 10^2$, leading to
$\beta = 10^{-2}$ and $D \sim 10^{-4}$. As noted above, however, this
estimate of $D$ is a conservative one, and $f_i$ values higher by factors 
of three or more are likely possible. Hence we conclude that on the basis
of these 3D models, values of $D \sim 10^{-4}$ to $\sim 3 \times 10^{-4}$
are plausible.

 Takigawa et al. (2008) found that a dilution factor $D \sim 10^{-4}$ to
$\sim 10^{-3}$ was needed in order to account for deriving the solar system's 
SLRIs $^{26}$Al, $^{41}$Ca, $^{53}$Mn, and $^{60}$Fe from a type II SNe. They
considered detailed models of core-collapse SNe, with mixing and fallback
of the interior layers of SNe with masses of 20 to 40 $M_\odot$, finding
dilution factors ranging from $1.34 \times 10^{-4}$ to $1.9 \times 10^{-3}$,
depending on the mass of the particular SNe. While Takigawa et al. (2008) assumed 
the canonical initial $^{26}$Al/$^{27}$Al abundance of $5 \times 10^{-5}$
and an initial ratio of $^{41}$Ca/$^{40}$Ca = $1.4 \times 10^{-8}$, they also 
assumed an initial $^{60}$Fe/$^{56}$Fe ratio of $7.5 \times 10^{-7}$, based on 
the work of Tachibana et al. (2006), who reported an initial $^{60}$Fe/$^{56}$Fe
in the range of $5 - 10 \times 10^{-7}$. Tang \& Dauphas (2012), however, derived 
an initial $^{60}$Fe/$^{56}$Fe ratio of $1.1 \times 10^{-8}$, based on their
analyses of a wide range of meteorites, and concluded that such a low level
of $^{60}$Fe could be explained by the galactic background level of 
$^{60}$Fe/$^{56}$Fe $\sim 2.8 \times 10^{-7}$ inferred from $\gamma$-ray 
observations. On the other hand, Liu (2014) reconsidered whether a supernova with 
mixing and fallback might account for this lower initial $^{60}$Fe/$^{56}$Fe ratio,
as well as for the lower initial $^{41}$Ca/$^{40}$Ca = $4.2 \times 10^{-9}$ 
ratio found by Liu et al. (2012), and concluded that such a SNe scenario
was still possible. In addition, Mishra \& Goswami (2014) concluded from
their studies of $^{60}$Fe and $^{26}$Al systematics in primitive chondrites
that the initial $^{60}$Fe/$^{56}$Fe ratio was $7.0 \pm 1.2 \times 10^{7}$,
essentially the same as the ratio assumed by Takigawa et al. (2008). 

 While the situation regarding the initial ratio of $^{60}$Fe is probably still 
in flux, on the basis of the most recent work, the dilution factors estimated
by Takigawa et al. (2008) to be in the range of $D \sim 10^{-4}$ to $\sim 10^{-3}$ 
appear to be reasonable for the solar system's SLRIs being derived primarily 
from a core-collapse SNe. Given that the present 3D AMR models suggest dilution 
factors in the range $D \sim 10^{-4}$ to $\sim 3 \times 10^{-4}$, these models 
appear to be consistent with the initial ratios inferred for these SLRIs.

 Supernova SLRIs are known to be ejected highly anisotropically, perhaps indicative 
of convective instabilities in core-collapse supernovae (Grefenstette et al. 2014).
Isotopes synthesized in different layers of the SNe have quite different, clumpy spatial 
distributions even shortly after the explosion, e.g., 340 yr after the formation of the 
Cassiopeia A SNR, when it has a radius of only $\sim$ 1.8 pc (Grefenstette et al. 2014). 
This implies that the abundances of different SLRIs will vary strongly across a SNR, 
and by the time that a SNR expands to a radius of $\sim$ 10 pc, sweeping up 
intervening ISM gas and slowing down enough to trigger the collapse of dense 
cloud cores (as in the W44 SNR; Sashida et al. 2013), the SLRI abundances
in the portion of a SNR that is injected into a particular collapsing cloud core 
may be quite different from what would be expected if all the SLRIs had been ejected
in a highly isotropic manner. Hence the actual abundances injected could be 
significantly higher, or lower, by factors of $\sim 4$ (based on the error bars
on the total yield of $^{44}$Ti for the Cass A SNR of $1.25 \pm 0.3 \times 10^{-4}$ 
M$_\odot$ found by Grefenstette et al. 2014, and the fact that some regions showed no 
evidence for $^{44}$Ti at the error bar level) than predicted abundances based 
on isotropic SNe explosions. This factor of $\sim$ 4 appears to be more than enough 
to account for any remaining discrepancies between the results of these 3D AMR models 
and the ongoing laboratory estimates of the initial abundances of SLRIs.

\section{Conclusions}

 These detailed 3D AMR hydrodynamics calculations have shown that SNe shock
waves can trigger the collapse of rotating dense cloud cores, resulting in
dynamic collapse leading to the formation of single, central protostars
embedded in protostellar disks. Such systems are likely to evolve into 
solar-type protostars with protoplanetary disks similar to the solar nebula.
Shock wave material carrying SLRIs produced by the SNe is injected into the
collapsing regions through Rayleigh-Taylor fingers, leading to an initially
inhomogeneous distribution, with SLRI abundances higher in the regions outside 
the densest regions of the protostellar disks, and lower abundances in the 
central protostar. This result is qualitatively consistent with the apparent need 
for late injection of SLRIs into the solar nebula (e.g., Sahijpal \& Goswami 1998; 
Krot et al. 2012; Kita et al. 2013; Mishra \& Chaussidon 2014). The estimated 3D
model dilution factors for a SNR from a core-collapse SNe are in the range of
$\sim 10^{-4}$ to $\sim 3 \times 10^{-4}$, in agreement with recent 
laboratory estimates, which require dilution factors in the range of
$\sim 10^{-4}$ to $\sim 10^{-3}$, depending on the mass of the SNe. The
significant anisotropy of isotopes observed in SNRs such as Cass A suggests 
that this level of agreement may be the best that can be expected, i.e.,
even this order of magnitude agreement should be considered a success.

 Our future 3D FLASH models will investigate the outcome of shock-cloud
interactions where the target cloud's rotation axis is {\it perpendicular}
to the direction of shock propagation, instead of aligned, as in
the present set of models, to learn what effect such orientations might
have on injection efficiencies. We also plan to include the loss of
molecular line cooling once the clouds become optically thick in our
future models, i.e., at densities above $\sim 10^{-13}$ g cm$^{-3}$,
which will allow the collapsing regions to heat above 10 K and continue
their collapse toward the formation of the first protostellar core, at
$\rho_{max} \sim 10^{-10}$ g cm$^{-3}$ (e.g., Boss \& Yorke 1995).

\acknowledgments

We thank the referee for a number of suggestions for improving the 
manuscript, and Michael Acierno for his help with the flash cluster
at DTM, where the calculations were performed.
This research was supported in part by NASA Origins of Solar Systems 
grant NNX09AF62G and is contributed in part to NASA Astrobiology 
Institute grant NNA09DA81A. The software used in this work was in 
part developed by the DOE-supported ASC/Alliances Center for 
Astrophysical Thermonuclear Flashes at the University of Chicago.

\clearpage
\begin{deluxetable}{lccccccc}
\tablecaption{Initial conditions and results for the 3D models, with varied
shock speeds ($v_s$, in units of km s$^{-1}$), shock gas densities ($\rho_s$, 
in units of the standard shock density, $3.6 \times 10^{-20}$ g cm$^{-3}$),
shock widths ($w_s$, in units of the standard shock width, 0.0030 pc), and
target cloud solid body rotation rates ($\Omega_c$, in rad s$^{-1}$). Also
listed are the outcomes, and when available for the models where collapse to 
form a single protostar and disk occurred, the amounts of mass in the collapsing 
regions ($\rho > 10^{-16}$ g cm$^{-3}$) derived from the shock wave ($M_s$, in 
units of $M_\odot$) and the injection efficiencies ($f_i$) for the shock wave 
material. \label{tbl-1}}
\tablewidth{0pt}
\tablehead{\colhead{Model} 
& \colhead{$v_s$}
& \colhead{$\rho_s$} 
& \colhead{$w_s$} 
& \colhead{$\Omega_c$} 
& \colhead{outcome}
& \colhead{$M_s$} 
& \colhead{$f_i$} }
\startdata
20-10-1-13 & 20 & 10 & 1 & $10^{-13}$ & collapses to multiple &  ---  &  ---   \\

20-10-1-14 & 20 & 10 & 1 & $10^{-14}$ & collapses to single   &  0.030 &  0.11 \\ 

20-200-0.1-13 & 20 & 200 & 0.1 & $10^{-13}$ & collapses to multiple &  ---  &  ---   \\

20-200-0.1-14 & 20 & 200 & 0.1 & $10^{-14}$ & collapses to single   & 0.030 & 0.097 \\ 

20-400-0.1-13 & 20 & 400 & 0.1 & $10^{-13}$ & collapses to multiple &  ---  &  ---   \\ 

20-400-0.1-14 & 20 & 400 & 0.1 & $10^{-14}$ & collapses to single   &  ---  &  ---   \\ 

20-600-0.1-13 & 20 & 600 & 0.1 & $10^{-13}$ & collapses to multiple &  ---  &  ---   \\

20-600-0.1-14 & 20 & 600 & 0.1 & $10^{-14}$ & collapses to single   & 0.036 & 0.038 \\ 

40-10-1-13 & 40 & 10 & 1 & $10^{-13}$ & collapses to single   & 0.0088 & 0.034 \\ 

40-10-1-14 & 40 & 10 & 1 & $10^{-14}$ & collapses to single   & 0.0094 & 0.036 \\ 

40-200-0.1-13 & 40 & 200 & 0.1 & $10^{-13}$ & collapses to single   &  0.010 & 0.032 \\ 

40-200-0.1-14 & 40 & 200 & 0.1 & $10^{-14}$ & collapses to single   &  0.011 & 0.034 \\ 

40-400-0.1-13 & 40 & 400 & 0.1 & $10^{-13}$ & collapses to single   &  ---   & ---   \\ 

40-400-0.1-14 & 40 & 400 & 0.1 & $10^{-14}$ & collapses to single   &  ---   & ---   \\ 

40-600-0.1-13 & 40 & 600 & 0.1 & $10^{-13}$ & no sustained collapse &  ---   & ---   \\ 

40-600-0.1-14 & 40 & 600 & 0.1 & $10^{-14}$ & collapses to single   & 0.0052 & 0.0055 \\ 

60-200-0.1-13 & 60 & 200 & 0.1 & $10^{-13}$ & collapses to single   &  ---   & ---    \\ 

60-200-0.1-14 & 60 & 200 & 0.1 & $10^{-14}$ & collapses to single   & 0.0054 & 0.018  \\ 

60-400-0.1-13 & 60 & 400 & 0.1 & $10^{-13}$ & no sustained collapse &  ---   & ---    \\ 

60-400-0.1-14 & 60 & 400 & 0.1 & $10^{-14}$ & no sustained collapse &  ---   & ---    \\ 
\enddata
\end{deluxetable}

\begin{figure}
\vspace{-0.5in}
\plotone{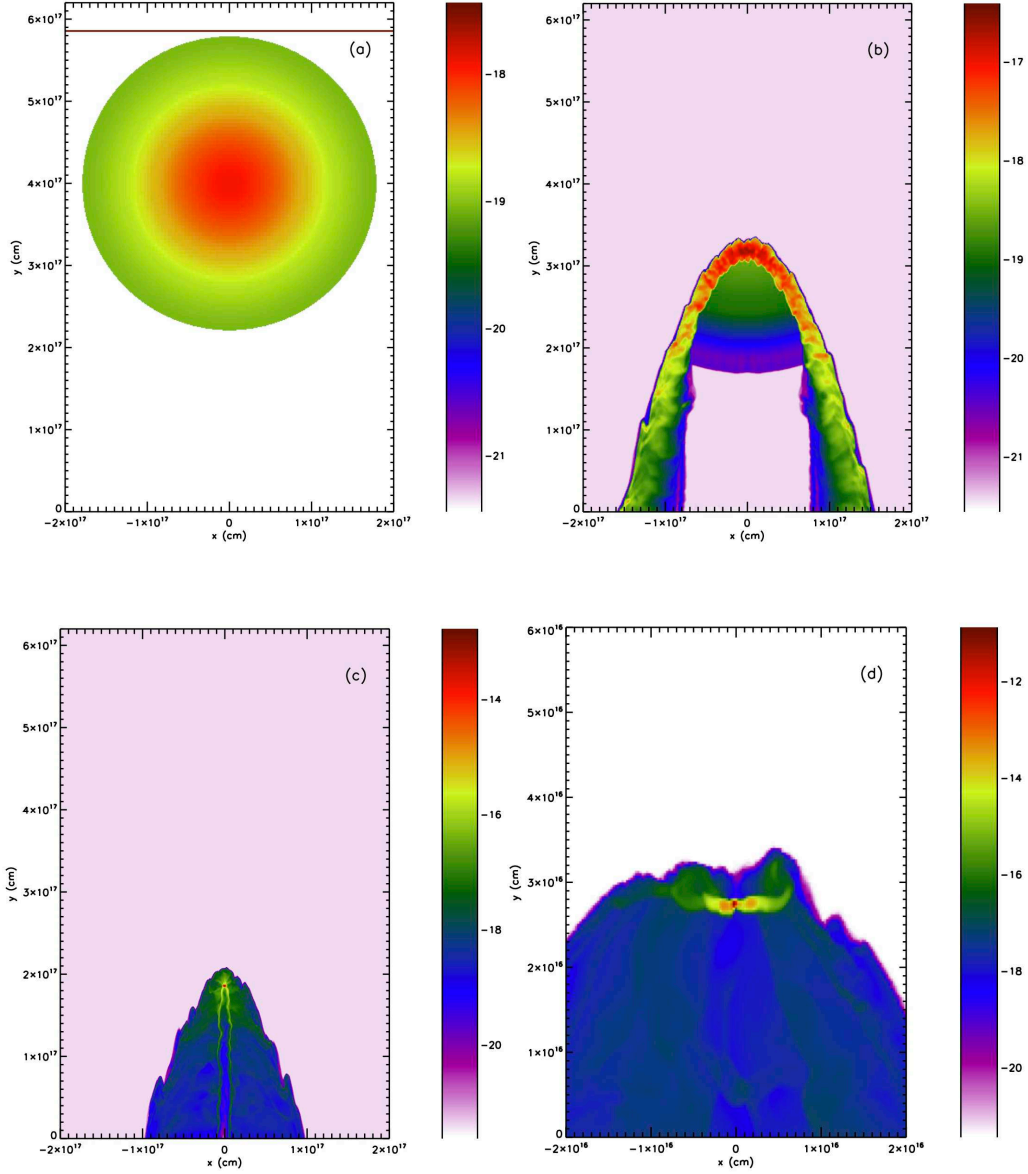}
\vspace{-1.5in}
\caption{Time evolution of model 40-200-0.1-13, showing the log10 of the density in a 
cross-section along the rotation axis. The initial shock front and target cloud are
shown in (a), and the Rayleigh-Taylor fingers responsible for injecting shock wave
material into the collapsing cloud are seen in (b). The formation of an edge-on disk 
is evident in (c) and (d), the latter of which is expanded by a factor of 10 for clarity. 
Times shown are: (a) 0.0 Myr, (b) 0.0280 Myr, (c) 0.0539 Myr, and (d) 0.0833 Myr.}
\end{figure}
\clearpage

\begin{figure}
\vspace{-1.0in}
\plotone{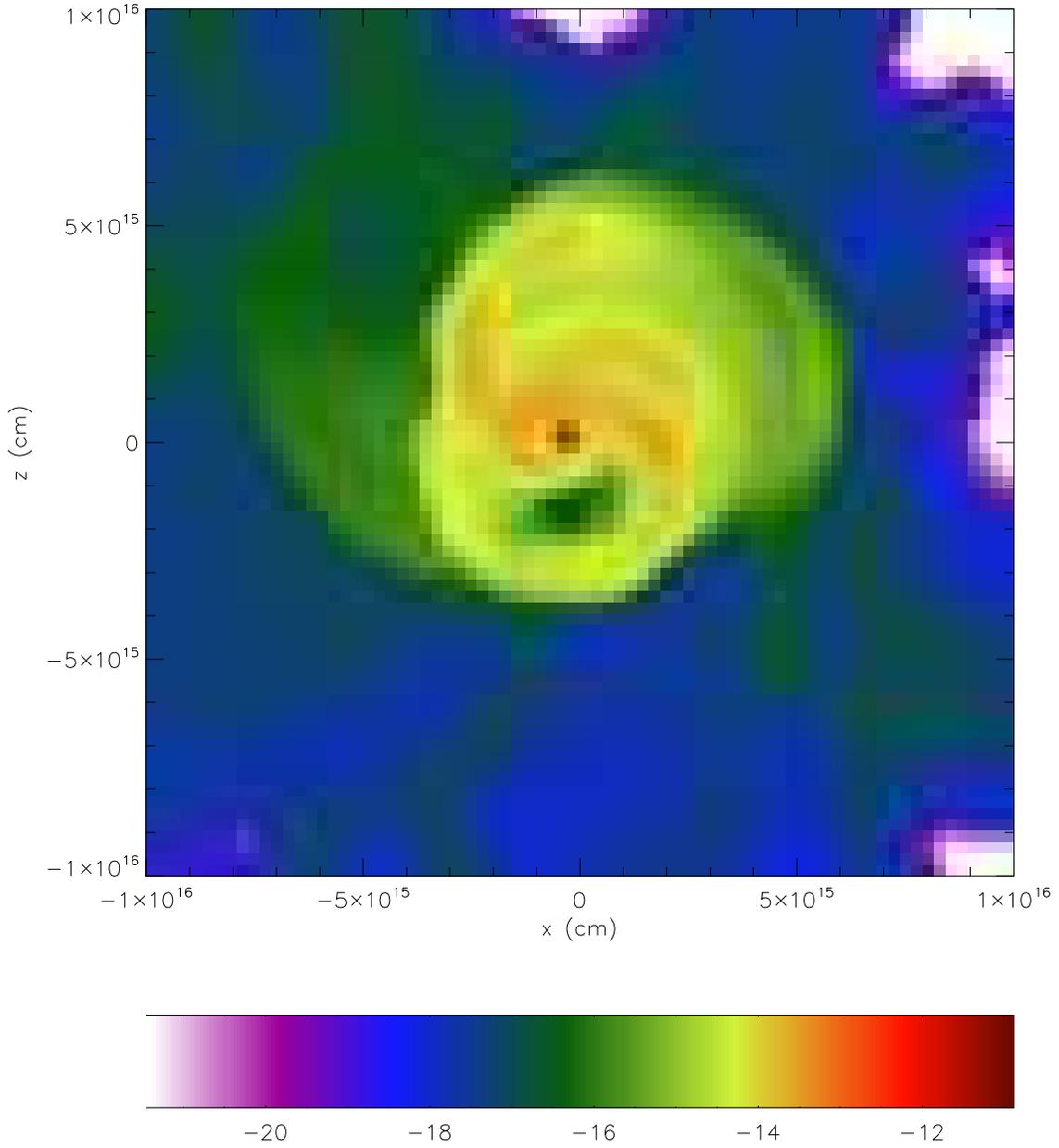}
\vspace{-0.5in}
\caption{Model 40-200-0.1-13 log10 density cross-section perpendicular to the rotation 
axis, showing the disk's midplane at $y = 2.75 \times 10^{16}$ cm in Fig. 1d. The
region is shown at 0.0833 Myr and several distinct spiral arms are evident, in a
disk with a radius of $\sim$ 500 AU.}
\end{figure}
\clearpage

\begin{figure}
\vspace{-1.0in}
\plotone{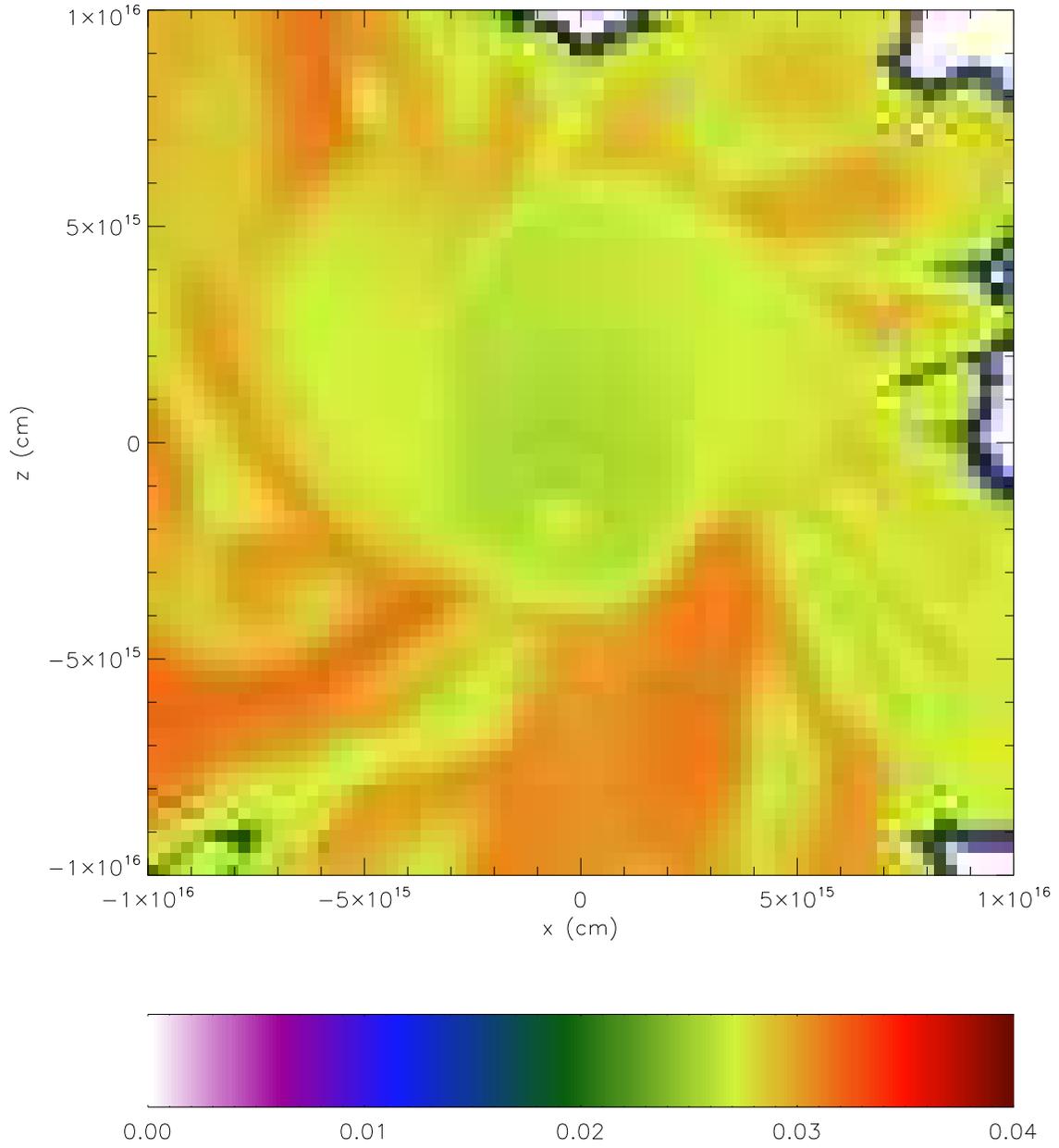}
\vspace{-0.5in}
\caption{Model 40-200-0.1-13 shock wave material (color field) in the disk's midplane at 
0.0833 Myr, plotted as in Fig. 2. Shock wave SLRIs have been injected throughout 
this region, with higher concentrations (orange) surrounding the innermost disk 
(light green).}
\end{figure}
\clearpage

\begin{figure}
\vspace{-0.0in}
\plotone{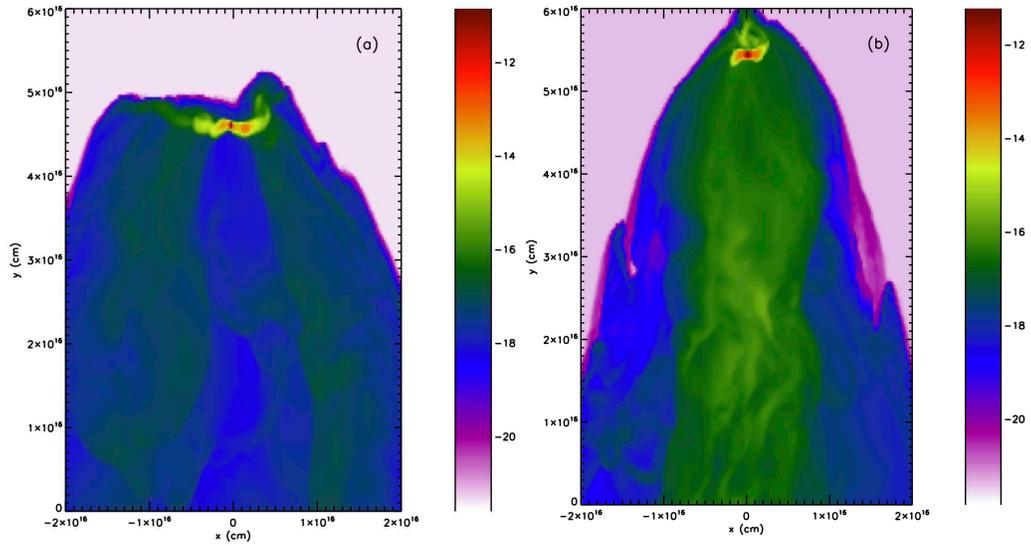}
\vspace{-4.5in}
\caption{Edge-on view of the disks formed by models (a) 40-200-0.1-13 and
(b) 40-200-0.1-14, differing only in the initial cloud rotation rate,
plotted as in Fig. 1, at similar times: (a) 0.0799 Myr and (b) 0.0779 Myr.
The disk radius for model 40-200-0.1-13 is $\sim$ 500 AU, while that for
model 40-200-0.1-14 is $\sim$ 150 AU.}
\end{figure}
\clearpage

\begin{figure}
\vspace{-1.0in}
\plotone{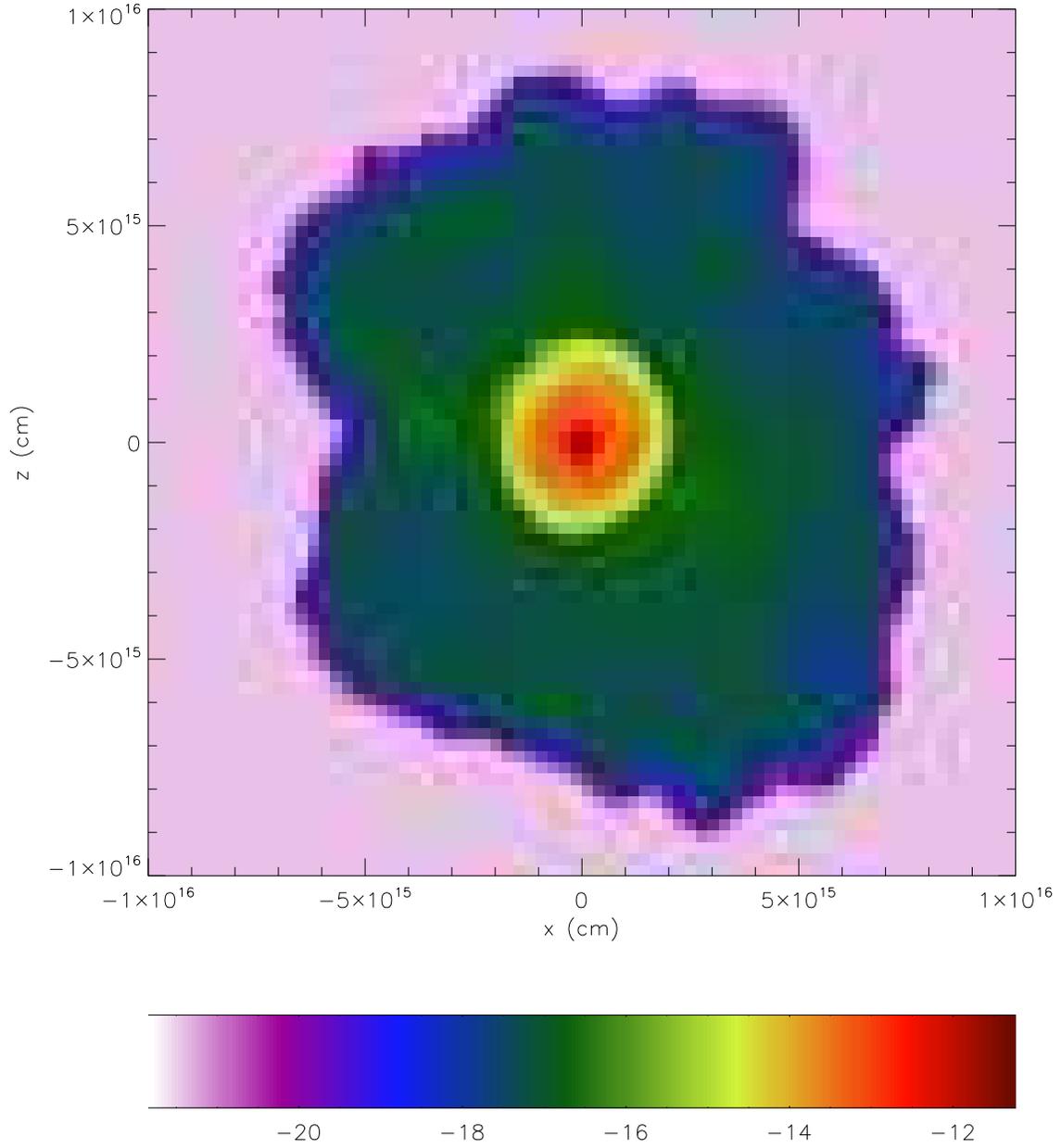}
\vspace{-0.5in}
\caption{Model 40-200-0.1-14 density at 0.0779 Myr, plotted as in Fig. 2,
showing the disk's midplane at $y = 5.40 \times 10^{16}$ cm in Fig. 4b.
Compared to Fig. 2, no spiral arms are evident in this considerably
smaller disk at this phase of evolution.}
\end{figure}
\clearpage

\begin{figure}
\vspace{-1.0in}
\plotone{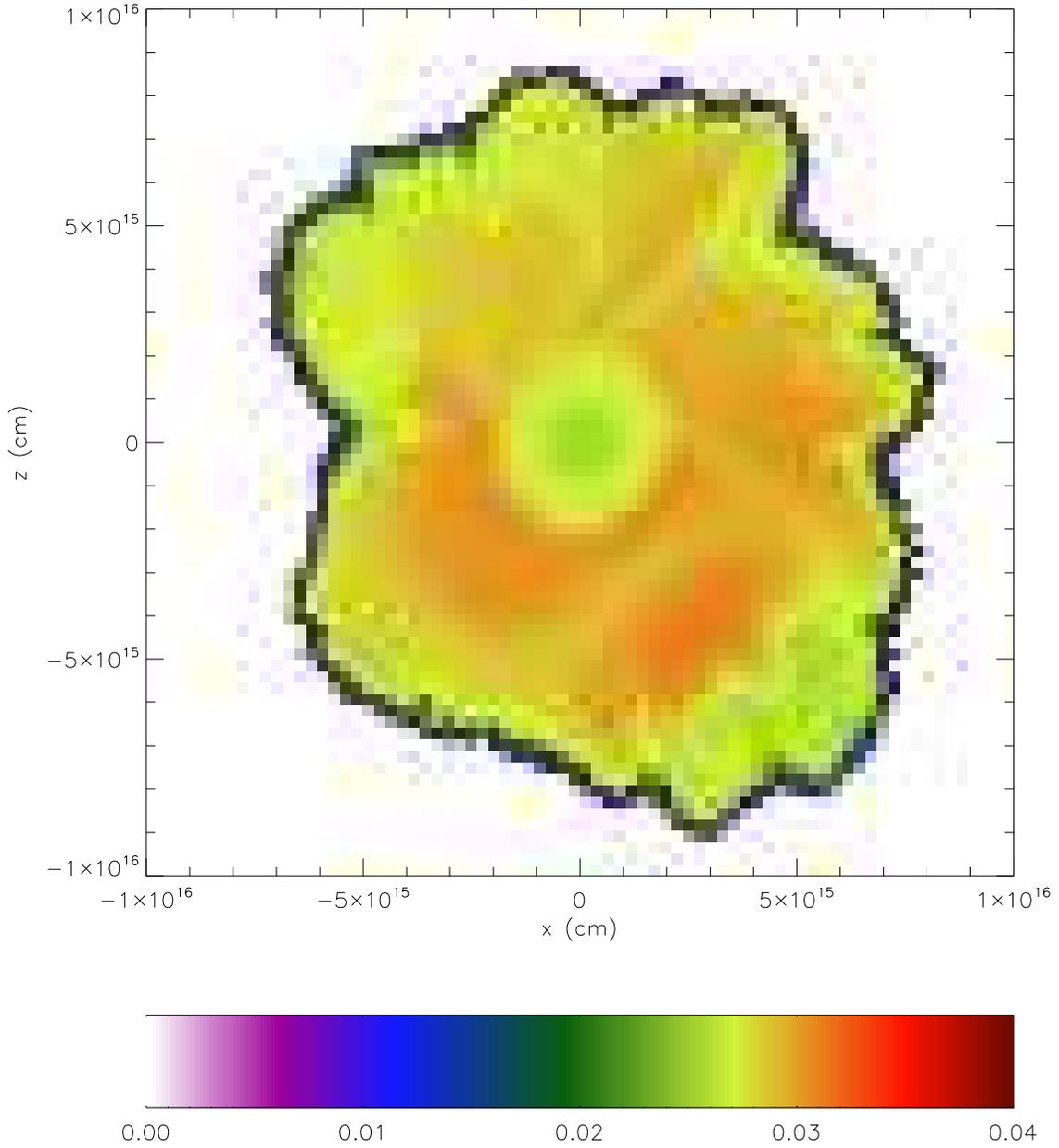}
\vspace{-0.5in}
\caption{Model 40-200-0.1-14 shock wave material (color field) at 0.0779 Myr, 
plotted as in Fig. 5, showing the SLRI abundances in the disk's midplane, 
with higher SLRI abundances (orange) in the outermost regions shown. Vestigal 
Rayleigh-Taylor fingers are evident as these orange clumps.}
\end{figure}
\clearpage

\begin{figure}
\vspace{-1.0in}
\plotone{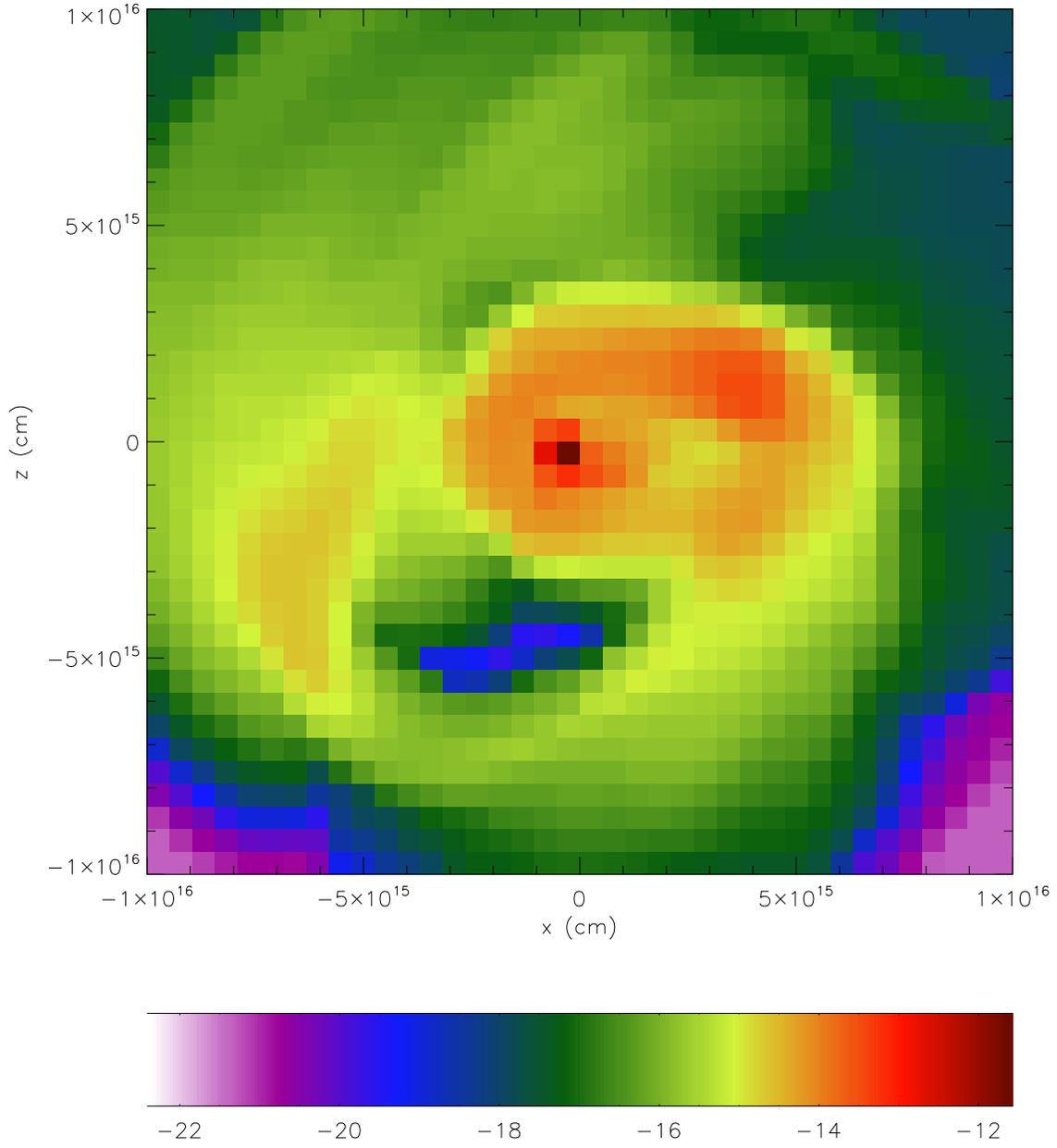}
\vspace{-0.5in}
\caption{Model 20-200-0.1-13 density at 0.0955 Myr, plotted as in Fig. 2,
showing the disk's midplane at $y = 1.70 \times 10^{17}$ cm. A multiple
protostar system has formed.}
\end{figure}
\clearpage

\begin{figure}
\vspace{-1.0in}
\plotone{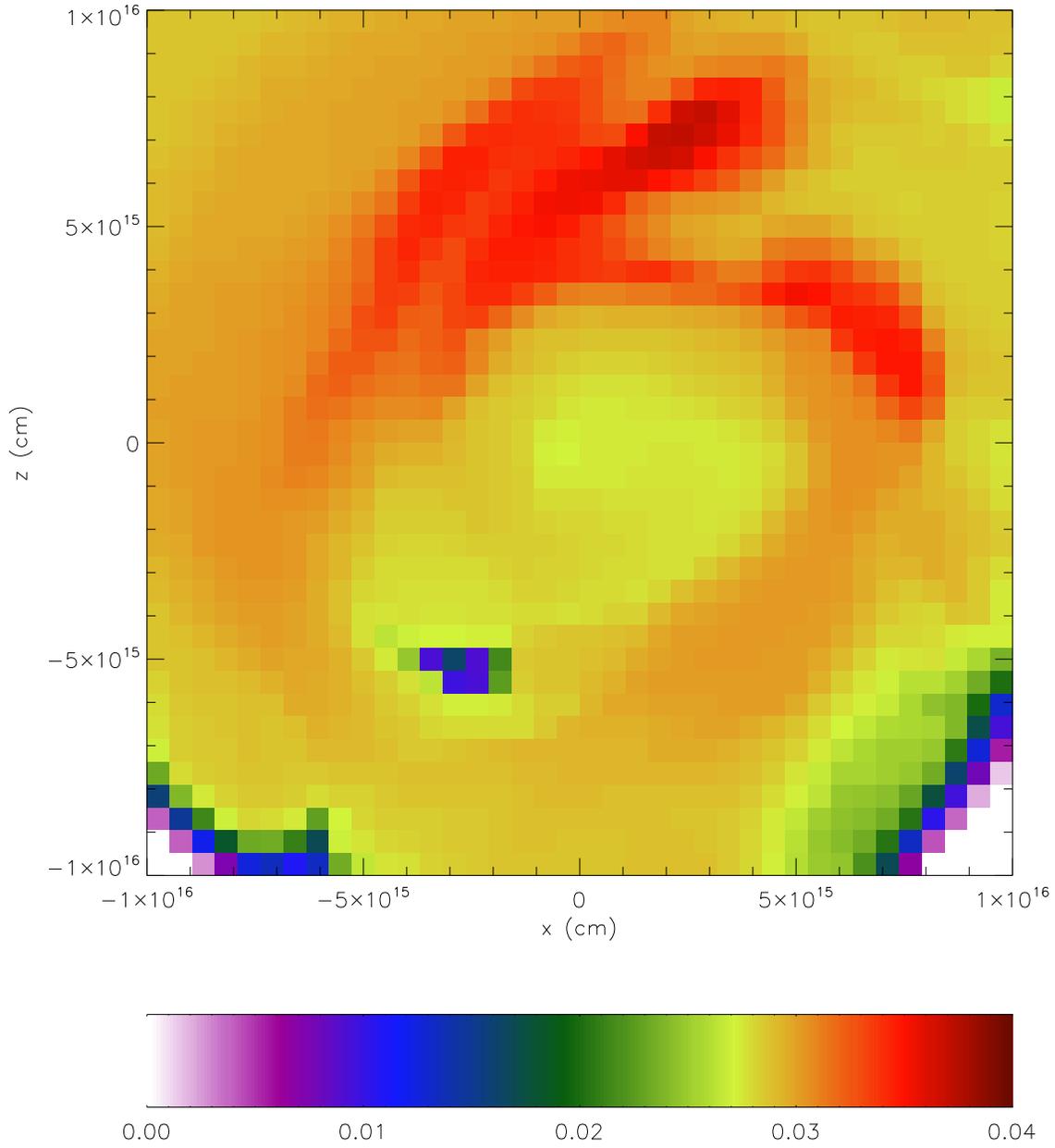}
\vspace{-0.5in}
\caption{Model 20-200-0.1-13 shock wave material at 0.0955 Myr, plotted as 
in Fig. 5, showing the disk's midplane at $y = 1.70 \times 10^{17}$ cm. 
Strong gradients in the SLRI abundances are evident.}
\end{figure}
\clearpage

\end{document}